\documentclass{amsart}

\makeatletter

\newtheorem{thm}{Theorem}[section]
\def\statetheorem{\@ifnextchar[{\@statetheorem}{\nr@statetheorem}}
\long\def\@statetheorem[#1]#2{\begin{thm}\label{#1}#2\end{thm}}
\long\def\nr@statetheorem#1{\begin{thm}#1\end{thm}}
\def\statetheorempf{\@ifnextchar[{\@statetheorempf}{\nr@statetheorempf}}
\long\def\@statetheorempf[#1]#2{\begin{thm}\label{#1}#2\end{thm}\proof}
\long\def\nr@statetheorempf#1{\begin{thm}#1\end{thm}\proof}

\newtheorem{lmma}{Lemma}[section]
\def\statelemma{\@ifnextchar[{\@statelemma}{\nr@statelemma}}
\long\def\@statelemma[#1]#2{\begin{lmma}\label{#1}#2\end{lmma}}
\long\def\nr@statelemma#1{\begin{lmma}#1\end{lmma}}
\def\statelemmapf{\@ifnextchar[{\@statelemmapf}{\nr@statelemmapf}}
\long\def\@statelemmapf[#1]#2{\begin{lmma}\label{#1}#2\end{lmma}\proof}
\long\def\nr@statelemmapf#1{\begin{lmma}#1\end{lmma}\proof}

\newtheorem{crlry}{Corollary}[section]
\def\statecorollary{\@ifnextchar[{\@statecorollary}{\nr@statecorollary}}
\long\def\@statecorollary[#1]#2{\begin{crlry}\label{#1}#2\end{crlry}}
\long\def\nr@statecorollary#1{\begin{crlry}#1\end{crlry}}
\def\statecorollarypf{\@ifnextchar[{\@statecorollarypf}{\nr@statecorollarypf}}
\long\def\@statecorollarypf[#1]#2{\begin{crlry}\label{#1}#2\end{crlry}\proof}
\long\def\nr@statecorollarypf#1{\begin{crlry}#1\end{crlry}\proof}

\def\ord{\textup{ord}\,}

\def\C{\mathbb{C}}
\def\N{\mathbb{N}}
\def\bibref[#1]{\cite{#1}}

\let\real@bibitem\bibitem
\def\bibitem[#1]{\real@bibitem{#1}}

\makeatother

\title{Spectral Difference Equations Satisfied by KP Soliton Wavefunctions}
\author{Alex Kasman}
\address{Mathematical Sciences Research
Institute \\ Berkeley, CA 94720}

\def\genT{{\hat T}}

\def\hatLambda{{\hat\Lambda}}

\def\D{{\mathbb{D}}}

\def\TODOs{{\mathbb{T}}}
\def\L{{\mathcal L}}
\def\Trans{{\textbf{S}}}
\begin{document}


\begin{abstract}{It is by now well known that the wave functions of rational
solutions to the KP hierarchy which can be achieved as limits of the
pure $n$-soliton solutions satisfy an eigenvalue equation for ordinary
differential operators in the spectral parameter.  This property is
known as ``bispectrality'' and has proved to be both interesting and
useful.  In this note, it is shown that all pure soliton solutions of
the KP hierarchy (as well as their rational degenerations) satisfy an
eigenvalue equation for a non-local operator constructed by composing
ordinary differential operators in the spectral parameter with {\it
translation\/} operators in the spectral parameter, and therefore have
a form of bispectrality as well.}
\end{abstract}
\maketitle

\section{Introduction}
\subsection{The KP Hierarchy and Bispectrality}
Let $\D$ be the vector space spanned over $\C$ by the set
$$
\{\Delta(j,\lambda)\mid \lambda\in\C, j\in\N \}
$$
whose elements differentiate and evaluate functions of the variable
$z$:
$$
\Delta(j,\lambda)[f(z)]:=f^{(j)}(\lambda).
$$
The elements of $\D$ are thus \emph{finitely supported
  distributions} on appropriate spaces of functions in $z$.  For lack
of a better term, we will continue to call them distributions even
though their main use in this paper will be their application to
functions of two variables.    (Such
distributions were called ``conditions'' in \bibref[W] since a
KP wave function was specified by requiring that it be in their
kernel.)  Note that if $c\in\D$ and $f(x,z)$
is sufficiently differentiable in $z$ on the support of $c$, then
$\hat f(x)=c[f(x,z)]$ is a function of $x$ alone.  Furthermore,
note that one may ``compose'' a distribution with a function of $z$,
i.e. given $c\in\D$ and $f(z)$ (sufficiently differentiable on the
support of $c$) then there exists a $c':=c\circ f\in\D$ such
that
$$
c'(g(z))=c(f(z)g(z))\qquad\forall g.
$$

The subspaces of $\D$ can be used to generate solutions to the KP
hierarchy \bibref[SW] in the following way.  Let $C\subset\D$ be an
$n$ dimensional subspace with basis $\{c_1,\ldots,c_n\}$.  Then, if
$K=K_C$ is the unique, monic ordinary differential operator in $x$ of
order $n$ having the functions $c_i(e^{xz})$ in its kernel (see
\eqref{eqn:wrformK}) we define $\L_C=K \frac{\partial}{\partial x} K^{-1}$ and
$\psi_C=\frac{1}{z^n}K e^{xz}$.  The connection to integrable systems
comes from the fact that adding dependence to $C$ on a sequence of
variables $t_j$ ($j=1,2,\ldots$) by letting $C(t_j)$ be the space with
basis $$\{c_1\circ e^{\sum -t_jz^j},c_2\circ e^{\sum -t_jz^j},\ldots,
c_n\circ e^{\sum -t_jz^j}\}$$ it follows that the ``time dependent''
pseudo-differential operator $\L=\L(t_j)$ satisfies the equations of
the KP hierarchy \bibref[BHYsato,thesis,cmbis,W] $$
\frac{\partial}{\partial t_j}\L=[(\L^j)_+,\L].
$$
The {\it wave function\/}
$\psi_C(x,z)$ generates the corresponding subspace of the infinite
dimensional grassmannian $Gr$ \bibref[SW] which parametrizes KP solutions and
thus it is not difficult to see that this construction produces
precisely those solutions associated to the subgrassmannian
$Gr_1\subset Gr$ \bibref[SW,W].

Moreover, the ring $A_C=\{p\in\C[z]|c_i\circ p\in C\ 1\leq i \leq n\}$
is necessarily non-trivial (i.e.\ contains non-constant polynomials) and the operator $L_p=p(\L)$ is an {\it
ordinary\/} differential operator for every $p\in A_C$ and satisfies
\begin{equation}
L_p\psi_C(x,z)=p(z) \psi_C(x,z).\label{eigenx}
\end{equation}
The subject of this paper is the existence of {\it additional\/} eigenvalue
equations satisfied by $\psi_C(x,z)$.  In particular, we wish to
consider the question of whether there exists an
operator $\hatLambda $ acting on functions of the variable $z$ such that
\begin{equation}
\hatLambda \psi_C(x,z)=\pi(x)\psi_C(x,z)\label{eigenz}
\end{equation}
where $\pi(x)$ is a non-constant function of $x$.
For example, the following theorem is due to G. Wilson in \bibref[W]:
\statetheorem[Th:wilson]{In addition to \eqref{eigenx} the wave function $\psi_C(x,z)$ is also an
eigenfunction for a ring of ordinary differential operators in $z$
with eigenvalues depending polynomially on $x$ if and only if $C$ has a
basis of distributions each of which is supported only at one point.}
In other words, for this special class of KP solutions for which the
coefficients of $\L$ are rational functions of $x$, the wave
function $\psi_C$ satisfies an additional eigenvalue equation of the form
\eqref{eigenz}
where $\hatLambda$ is an ordinary differential operator in $z$ and
$\pi(x)$ a non-constant polynomial in $x$.\footnote{Moreover, he
demonstrated that up to conjugation or change of variables, the
operators $L_p$ found in this way are the only bispectral operators
which commute with differential operators of relatively prime order,
but this fact will not play an important role in the present note.}
Together \eqref{eigenx} and \eqref{eigenz} are an example of {\it
bispectrality\/} \bibref[DG,G].  The bispectral property is already
known to be connected to other questions of physical significance such
as the time-band limiting problem in tomography \bibref[Gr1], Huygens'
principle of wave propagation \bibref[Yuri], quantum integrability
\bibref[HK,Vbis] and, especially in the case described above, the self
duality of the Calogero-Moser particle system \bibref[cmbis,W,W2].

It is known that the only subspaces $C$ for which the corresponding
wave function satisfies \eqref{eigenx} and \eqref{eigenz} with $L_p$
and $\hatLambda $ ordinary differential operators in $x$ and $z$
respectively are those described in Theorem~\ref{Th:wilson}.  However,
suppose we allow $\hatLambda$ to involve not only differentiation and
multiplication in $z$ but also {\it translation\/} in $z$ and call
this more {\it general\/} situation t-bispectrality.\footnote{It
should be noted that the term ``bispectrality'' already applies to
more general situations than simply differential operators \bibref[G],
but in the case of the KP hierarchy I believe only differential
bispectrality has thus far been considered.}  It will be shown below
that there are more KP solutions which are bispectral in this sense.
In particular, it will be shown that the KP solution associated to
\textit{any} subspace $C$ shares its eigenfunction with a ring of
translational-differential operators in the spectral parameter.

\subsection{Notation}


Using the shorthand notation 
$\partial=\frac{\partial}{\partial x}$ any ordinary differential
operator in $x$ can be
written as
$$
L=\sum_{i=0}^N f_i(x) \partial^i\qquad (N\in\N).
$$
We say that a function of the form
$$
f(x)=\sum_{i=1}^n p_i(x)e^{\lambda_i x}\qquad \lambda_i\in\C,\
p_i\in\C[x]
$$
is a \textit{polynomial-exponential function} and that the quotient of
two such functions is \textit{rational-exponential}.  
This note will be especially concerned with the ring of differential
operators with rational-exponential coefficients and especially with
the subring having polynomial-exponential coefficients.
Similarly, we will write
$\partial_z=\frac{\partial}{\partial z}$ but will need to consider
only differential operators in $z$ with rational coefficients.

For any $\lambda\in\C$ let $\Trans_{\lambda}=e^{\lambda\partial_z}$ be the
translational operator acting on functions of $z$ as
$$
\Trans_{\lambda}[f(z)]=f(z+\lambda).
$$
Then consider the ring of translational-differential operators
$\TODOs$ generated by these translational operators and ordinary
differential operators in $z$.  Any translational-differential operator $\hat
T\in\TODOs$  can be
written as
$$
\genT =\sum_{i=1}^N p_i(z,\partial_z) \Trans_{\lambda_i}
$$
where $p_i$ are ordinary differential operators in $z$ with rational
coefficients  and $N\in\N$.  Note that the ring of ordinary
differential operators in $z$ with rational coefficients is
simply the subring of $\TODOs$ of all elements which can be written as
$p \Trans_{0}$ for a differential operator $p$.

\section{Translational Bispectrality of $\C[\partial]$}

It has been frequently observed that the ring
$\mathcal{A}=\C[\partial]$ of constant coefficient differential
operators in $x$ is \textit{bispectral} since it has the eigenfunction
$e^{xz}$ which it shares with the ring of constant coefficient
differential operators in $z$.  Here, however, we will consider a more
general form of bispectrality for the ring $\mathcal{A}$.  

Let $\mathcal{A}'\subset\TODOs$ be the ring of constant
coefficicient \textit{translational}-differential operators.  Note
that for any element $\genT\in\mathcal{A}'$ of the form
$$\genT =\sum_{i=1}^N p_i(\partial_z) \Trans_{\lambda_i}$$
one has simply that
$$
\genT[e^{xz}]=\left(\sum_{i=1}^N p_i(x)e^{\lambda_i x}\right)e^{xz}.$$
In particular, $e^{xz}$ is an eigenfunction for the operator with an
eigenvalue which is a polynomial-exponential function of $x$.
Consequently, the rings $\mathcal{A}$ and $\mathcal{A}'$ are
both bispectral, sharing the common eigenfunction $e^{xz}$.

Let $\mathcal{R}$ be the ring of differential operators in $x$ with
polynomial-exponential coefficients and $\mathcal{R'}$ be the ring of
translational-differential operators in $z$ with rational
coefficients.  Note that $\mathcal{R}$ is generated by $\mathcal{A}$
and the eigenvalues of the operators in $\mathcal{A'}$ while
$\mathcal{R}'$ is generated by $\mathcal{A}'$ and the eigenvalues of
the elements of $\mathcal{A}$.  It then follows \bibref[BHYpla] (see also
\bibref[KR]) that the map $b:\mathcal{R}\to\mathcal{R}'$ defined by
the relationship
$$
L[e^{xz}]=b(L)[e^{xz}]\qquad \forall L\in\mathcal{R}
$$
is an anti-isomorphism of the two rings.

\section{Translational Bispectrality of KP Solitons}

Let us say that a finite dimensional
subspace $C\subset\D$ is {\it t-bispectral\/} if there exists a
translational-differential operator  $\hatLambda \in \TODOs$ satisfying
equation \eqref{eigenz} for the corresponding KP wave function
$\psi_C(x,z)$.
By Theorem~\ref{Th:wilson} and the fact that the ring of rational coefficient
ordinary differential operators in $z$ is contained in $\TODOs$, we
know that $C$ is t-bispectral\footnote{...and also bispectral
in the sense of \bibref[W].} if it has a basis of point supported
distributions.  Here we will show that, in fact, all subspaces
$C\subset\D$ are $t$-bispectral.

An important object in much of the literature on integrable systems is
the ``tau function''.  The tau function of the KP solution associated
to $C$ can be written easily in terms of the basis $\{c_i\}$.  In
particular, define (cf.\ \bibref[W])
$$
\tau_C(x)=\textup{Wr}\left(c_1(e^{xz}),c_2(e^{xz}),\ldots,c_n(e^{xz})\right)
$$
to be the Wronskian determinant of the functions $c_i(e^{xz})$.
Similarly, there is a Wronskian formula for the coefficients of the
operator $K_C$ since its action on an arbitrary function $f(x)$ is
given as:
\begin{equation}
K_C(f(x))=\frac{1}{\tau_C(x)} \textup{Wr}\left(c_1(e^{xz}),c_2(e^{xz}),\ldots,c_n(e^{xz}),f(x)\right).\label{eqn:wrformK}
\end{equation}
Then the coefficients of the differential operator $\bar
K_C:=\tau_C(x)K_C(x,\partial)$ are all polynomials-exponential
functions.

\begin{lemma}[lem:factor]
{For any $C\subset\D$ there exists a constant coefficient operator
$L_0\in\mathcal{A}$ which factors as
$$
L_0=\bar Q_g\circ\frac{1}{\pi(x)}\circ \bar K_C
$$
where $\bar Q_g,\bar K_C\in\mathcal{R}$ and $\pi(x)=g(x)\tau_C(x)\in\mathcal{R}$ is a
polynomial-exponential function.}
Let $\lambda_i\in\C$ ($1\leq i \leq N$) be the support of the
distributions in $C$ and $m_i$ be the highest derivative taken at
$\lambda_i$ by any element of $C$.  Then the polynomial
\begin{equation}
q_C(z):=(z-\lambda_i)^{m_i+1}\label{eqn:q}
\end{equation}
 has the property that $c\circ q_C\equiv0$
for any $c\in C$.  Let $L_0:=q_C(\partial)$ and consider
$L_0[c(e^{xz})]$ for any element $c\in C$.  Since $L_0$ is an operator
in $x$ alone, it commutes with $c$ and we have
$$
L_0[c(e^{xz})]=c\left(L_0[e^{xz}]\right)=c\left(q(z)e^{xz}\right)=c\circ
q(e^{xz})=0.
$$
So, by the definition of $K_C$, we see that $L_0$ annihilates the
kernel of $K_C$ and thus has a right factor of $K_C$.  This gives a
factorization of the form $L_0=Q\circ K_C$ with $Q$ having
rational-exponential coefficients.  Then, by choosing a
polynomial-exponential function $g(x)$ so that $\bar Q_g:=Q\circ
g(x)\in\mathcal{R}$ we find the desired factorization.
\end{lemma}

Given this factorization, the t-bispectrality of all $C$'s now follows
from Theorem 4.2 in \bibref[BHYpla]. 
\begin{theorem}[thm:main]
{For any subspace $C\subset\D$ the
corresponding KP wave function $\psi_C(x,z)$ satisfies the
eigenvalue equation
$$
\hatLambda_g[\psi_C(x,z)]=g(x)\tau_C(x)\psi_C(x,z)
$$
where $\hatLambda_g\in\TODOs$ is the 
translational-differential operator 
defined by $$\hatLambda_g:=z^{-n}\circ b(\bar K_C)\circ b(\bar Q_g)\circ \frac{z^n}{q_C(z)}
$$
with $\bar Q_g$ defined as in Lemma~\ref{lem:factor}.}
Formally introducing inverses \bibref[BHYpla], we determine from
Lemma~\ref{lem:factor} that
$$
\pi(x):=g(x)\tau_C(x)=\bar K_C \circ L_0^{-1} \circ \bar Q
$$
and hence (by applying the anti-involution $b$ to this equation)
$$
b(\pi(x))=b(\bar Q)\circ \frac{1}{q_C(z)}\circ b(\bar K_C).
$$
Then 
\begin{eqnarray*}
\hatLambda_g[\psi_C(x,z)] &=& 
z^{-n}\circ b(\bar K_C)\circ b(\bar Q)\circ
\frac{z^n}{q_C(z)}[\frac{1}{z^n \tau_C(x)} \bar K_C e^{xz}]\\
 &=& \frac{z^{-n}}{\tau_C(x)} \circ b(\bar K_C)\circ b(\bar Q)\circ
\frac{1}{q_C(z)}[\bar K_C e^{xz}]\\
 &=& \frac{z^{-n}}{\tau_C(x)} \circ b(\bar K_C)[\pi(x)e^{xz}]\\
 &=& \frac{z^{-n}\pi(x)}{\tau_C(x)} \circ \bar K_C[e^{xz}]\\
 &=& \pi(x)\psi_C(x,z)
\end{eqnarray*}
\end{theorem}

Note that according to Theorem~\ref{thm:main}, each operator
$\hatLambda_g$ satisfies an \textit{intertwining relationship} 
$$
W \circ b(\pi(x)) = \hatLambda_g \circ W
$$
with the constant coefficient operator $b(\pi(x))$ where
$W=z^{-n}\circ b(\bar K_C)$.  As a result we find that:

\statecorollary{The set of all such operators $\hatLambda_g$ for a
given subspace $C\subset\D$ form a commutative ring of
translational-differential operators.}

\section{Examples}

If we choose $C$ to be the two dimensional space spanned by $c_1=\Delta(1,0)$ and $c_2=\Delta(1,1)$ (a ``two-particle''
Calogero-Moser type solution) then
$$
\psi_C(x,z)=(1+\frac{2+x-(2x+x^2)z}{x^2z^2})e^{xz}.
$$
In this case the translational-differential operators $\hatLambda$
given by Theorem~\ref{thm:main} are simply ordinary differential
operator.  For instance,
$$
\hatLambda =\partial_z^3+\frac{3}{z-z^2}\partial_z^2-\frac{6z^2-12z+3}{z^3(z-1)^2}\partial
+ \frac{12z-6}{z^2(z-1)^2}$$
which satisfies $\hatLambda \psi_C(x,z)=x^3\psi_C(x,z)$ (as we would
expect from earlier results on bispectrality.)

However, if we had chosen instead $c_1=\Delta(0,1)+\Delta(0,-1)$ and
$c_2=\Delta(0,2)+\Delta(0,0)$ we would instead have 
the case of a 2-soliton solution with
$$
\psi_C(x,z)=(1-\frac{6+(3z-2)e^{2x}+2z-ze^{-2x}}{(e^x+e^{-x})^2z^2})e^{xz}.
$$
One finds from the procedure given in the theorem that 
\begin{eqnarray*}
\hatLambda   &=&  z^{-2}\circ\left( 
\left( 20z + 11{z^2} - 8{z^3} + 
     {z^4} \right)\Trans_{-3}
  + 
   \left( 60 - 68z - {z^2} + 8{z^3} + {z^4} \right) \Trans_{5}
  \right.\\
&& + 
  \left( -36 + 24z + 16{z^2} - 16{z^3} + 
     4{z^4} \right)\Trans_{-1}
 + 
   \left( -44 - 88z - 8{z^2} + 16{z^3} + 4{z^4} \right) \Trans_{3}\\
&&
\left.\left( -12 - 16z - 2{z^2} + 
     6{z^4} \right)\Trans_{1}
\right)\circ\frac{z^2}{z^4-2z^3-z^2+2z}
\end{eqnarray*}
satisfies $\hatLambda [\psi_C(x,z)]=e^{-3x}(1+e^{2x})^4\psi_C(x,z)$.

\section{Conclusions}

In addition to being a generalization of the results of \bibref[DG,W]
on bispectral ordinary differential operators, the present note may be
seen as a generalization of \bibref[Reach] in which wave functions of
$n$-soliton solutions of the KdV equation are shown to satisfy
difference equations in the spectral parameter. 
The idea that KP solitons might be translationally bispectral was
proposed in \bibref[KPsol1].

As in \bibref[DG,W], the equations \eqref{eigenx} and \eqref{eigenz}
lead to the well known ``ad'' relations associated to bispectral
pairs.
That is, defining the ordinary differential operator $A_m$ in $x$ and
the translational-differential operator $\hat A_m$ in $z$ by 
$$A_m=\textup{ad}_{L_p}^m(\pi(x))
\qquad
\hat A_m=(-1)^m\textup{ad}_{p(z)}^m(\hatLambda)
$$
one finds that $A_m\psi_C(x,z)=\hat A_m\psi_C(x,z)$.  Similarly, if
$$B_m=\textup{ad}_{\pi(x)}^m(L_p)
\qquad
\hat B_m=(-1)^m\textup{ad}_{\hatLambda}^m(p(z))
$$
then $B_m\psi_C(x,z)=\hat B_m\psi_C(x,z)$.  Note that whenever the
order of $B_{m-1}=N>0$ the order of $B_m$ cannot be greater than
$N-1$.  So, the familiar result that $B_m\equiv0$ and $\hat B_m\equiv0$
for $m>\ord L_p$ holds, which is clearly a strong restriction on
the operator $\hatLambda$.  However, unlike the case of bispectral
ordinary differential operators, one cannot conclude that $A_m\equiv0$
for sufficiently large $m$ since the order of $\hat A_m$ may not be
reduced by increasing $m$.

The bispectrality of the rational KP solutions \bibref[W] has been
shown to have a dynamical significance.  In particular, it was shown
that the \textit{bispectral involution} is the linearizing map for the
classical Calogero-Moser particle system \bibref[cmbis,W,W2].
Moreover, other bispectral KP solutions have been found to have
similar properties \bibref[cmbis2,Roth].  This would seem to indicate
that it is likely that the bispectrality of KP solitons also has a
dynamical significance, as a map between the classical Ruijsenaars and
Sutherland systems (cf.\ \bibref[ruij]).  In fact, such a bispectral
relationship between the \textit{quantum} versions of these systems
has been recently found in \bibref[chal].  The dynamical significance
of these results will be considered in a separate paper.



\begin{thebibliography}{11}
\def\akbibitem#1{\bibitem[#1]}
\bibitem[BHYpla]{B. Bakalov, E. Horozov and M. Yakimov, ``General methods for constructing
    bispectral operators'', Phys. Lett. A \textbf{222} (1996), no.~1-2,
  59--66.}

\bibitem[BHYsato]{B. Bakalov, E. Horozov and M. Yakimov,
``B\"acklund-Darboux Transformations in Sato's Grassmannian'', {\it Serdica 
Math. Journal\/}, 22 (1996) pp. 571-588}

\bibitem[BHYcommrings]{B. Bakalov, E. Horozov and M. Yakimov,
``Commutative Rings of Bispectral Ordinary Differential Operators'',
to appear {\it Comm.Math.Phys\/}, (1998).}

\akbibitem{Yuri}{Yu. Berest, ``Huygens' principle and the bispectral
problem'',  CRM Proceedings and Lecture Notes, volume 14, American
Mathematical Society, Providence, RI, pp. 11-30}
\akbibitem{DG}{J.J. Duistermaat and F.A. Gr\"unbaum, ``Differential
equations in the spectral parameter'', Comm. Math. Phys. 103 (1986)
177-240}

\akbibitem{chal}{O.A. Chalykh, ``Duality of the generalized Calogero
and Ruijsenaars problems'' {\it Uspekhi Mat. Nauk\/} 52 (1997), no. 6(318), 191--192}

\akbibitem{Gr1}{F.A. Gr\"unbaum, ``Time-band limiting and the
bispectral problem'', Comm. Pure Appl.  Math 47 (1994) no3. 307--328.}

\akbibitem{G}{F.A. Gr\"unbaum, ``Bispectral Musings''
CRM Proceedings and Lecture Notes, volume 14, American
Mathematical Society, Providence, RI, 1988 pp. 31--46.}

\akbibitem{HK}{E. Horozov and A. Kasman, ``Darboux Transformations for
Bispectral Quantum Integrable Systems'', in preparation.}

\bibitem[thesis]{A. Kasman, ``Rank $r$ KP Solutions with Singular
Rational Spectral Curves'', Ph.D. thesis, Boston University (1995)}

\bibitem[cmbis]{A. Kasman, ``Bispectral KP Solutions and
Linearization of Calogero-Moser Particle Systems'', {\it
Communications in Mathematical Physics} 172 (1995) pp. 427-448}

\akbibitem{cmbis2}{A. Kasman, ``The Bispectral Involution as a
Linearizing Map'', to appear, Proceedings of Workshop on
Calogero-Moser-Sutherland Models, J.F. van Diejen and L. Vinet eds.,
Springer-Verlag (1998).}


\akbibitem{KR}{A. Kasman and M. Rothstein, ``Bispectral darboux
transformations: the generalized Airy case'', Physica D 102 (1997) p.
159-173.}

\akbibitem{KPsol1}{A. Kasman,
``Bispectrality and KP Solitons'', Preprint CRM-2533}


\akbibitem{Reach}{M. Reach
``Difference equations for $N$-soliton solutions to KdV''
{\it Phys. Lett. A\/} 129 (1988), no. 2, pp. 101--105}

\akbibitem{Roth}{M. Rothstein, ``Explicit Formulas for the Airy and
Bessel Bispectral Involutions in Terms of Calogero-Moser Pairs'', {\it
CRM Proc. and Lecture Notes\/}, Vol. 14, (1998) pp. 105--110}

\akbibitem{ruij}{S.N.M. Ruijsenaars, ``Action-Angle Maps and
Scaterring Theory for Some Finite-Dimensional Integrable Systems'',
{\it Commun. Math. Phys.\/} 115 (1988) pp. 127--165}

\akbibitem{SW}{G. Segal and G. Wilson, ``Loop groups and equations of
KdV type''  {\it Publications Mathematiques
No. 61 de l'Institut des Hautes Etudes Scientifiques\/} (1985)
pp. 5-65}

\akbibitem{Vbis}{A. P. Veselov, Baker-Akhiezer functions and the bispectral problem in many
dimensions, CRM Proceedings and Lecture Notes, volume 14, American
Mathematical Society, Providence, RI, pp.  
123--129.}
\akbibitem{W} G.~Wilson, \emph{Bispectral commutative ordinary
    differential operators}, J. Reine Angew. Math. \textbf{442}
  (1993), 177--204.

\akbibitem{W2}{ G.~Wilson, \emph{Collisions of Calogero-Moser
Particles and an Adelic Grassmannian}, {\it Inventiones Mathematicae\/} 
 133 (1998), no. 1, 1--41}

\end{thebibliography}
\end{document}